\documentclass[prb,twocolumn,showpacs]{revtex4}
\usepackage{graphicx}
\usepackage{bm}
\usepackage{amssymb,amsmath}
\usepackage[usenames]{color}
\usepackage{epsfig}
\usepackage{hyperref}
\usepackage{subfigure}
\usepackage[sans]{dsfont}
\hypersetup{colorlinks=true, citecolor=blue,
linkcolor=blue,urlcolor=blue }

\begin{document}

\title{Irrational anomalies in one-dimensional Anderson localization}
\author{Reza Sepehrinia}
\address{School of Physics, Institute for Research in Fundamental Sciences, IPM, 19395-5531 Tehran, Iran}
\begin{abstract}
We revisit the problem of one-dimensional Anderson localization, by
providing perturbative expression for Lyapunov exponent of Anderson
model with next-nearest-neighbor (nnn) hopping. By comparison with
exact numerical results, we discuss the range of validity of the
naive perturbation theory. The stability of band center anomaly is
examined against the introduction of nnn hopping. New anomalies of
Kappus-Wegner type emerge at nonuniversal values of wavelength when
hopping to second neighbor is allowed. It is shown that covariances
in the first order of perturbation theory, develop singularities at
these resonant energies which enable us to locate them.
\end{abstract}
 \maketitle

\section{Introduction}

Localization of noninteracting particles in one and
quasi-one-dimensional systems can be formulated using transfer
matrices. It is also mostly accepted that localization properties in
two and three dimensions can be deduced via transfer matrix method
combined with finite size scaling on quasi-one-dimensional
geometries. Therefor the problem reduces to calculating growth rates
[Lyapunov exponents (LEs)] of products of random matrices. There are
a few cases in which analytical expression for LEs can be obtained.
Most cases have to be treated numerically. But the limit of
arbitrarily weak disorder is not accessible even numerically because
the convergence slows down extremely. So it would be of crucial
importance to develop a perturbation theory in this limit.
Particularly, existence of delocalized states can be judged thereby.
Perturbative expansions for LEs are provided \cite{PRM} for some
class of random matrices of the form $\mathbf{T}=\mathbf{A}+
\epsilon \mathbf{B}$, where $\mathbf{A}$ is nonrandom matrix and
$\epsilon\mathbf{B}$ is small random part. They require the
eigenvalues of $\mathbf{A}$ to have different moduli. Many
interesting situations which appear in localization do not fulfill
this condition.

Using nondegenerate perturbation theory, Thouless \cite{Thouless}
obtained a weak disorder expansion of LE for one-dimensional
Anderson model. Later numerical results \cite{MacK81} showed $9\%$
increase in localization length at the center of band. This
discrepancy was resolved by Kappus and Wegner when they developed a
degenerate perturbation theory for the band center.\cite{Kappus} The
failure of nondegenerate perturbation theory at the middle of the
band is known as Kappus-Wegner (KW) anomaly. The anomaly is a
manifestation of spatial periodicity of the system. It is known as
commensurability effect between lattice constant and electron
wavelength. Description of this effect using phase formalism offers
an analogy in classical dynamics.\cite{Lambert84,Hayn88}

Apart from the mathematical subtlety, there are remarkable physical
consequences at this point. Conductance distribution at this anomaly
deviates \cite{Schomerus03} from predicted distribution (log-normal)
by single-parameter scaling (SPS) theory. The occurrence of anomaly
is accompanied with breakdown of reflection phase randomization
\cite{Lambert84,Titov05} as well, which also is of basic assumptions
of SPS theory.\cite{Anderson} Systematic treatment of the band
center anomaly as well as anomalies at the energies
$E=2t\cos(\pi\alpha)$ with $\alpha$ rational, is already
established.\cite{DG} Quite recently a classification of anomalies
is given \cite{Schulz} for $2\times2$ transfer matrices according to
which the anomaly in the band center is of second order. Another
study is done via calculation of participation ratio (instead of LE)
by means of field theoretic tools which provides full statistics of
wave function at the band center anomaly.\cite{Yudson} This leads
the authors to conjecture that there is a hidden symmetry
responsible for integrability of the problem at this spectral point.

\section{Model}

\begin{figure}[t]
\epsfxsize8truecm \epsffile{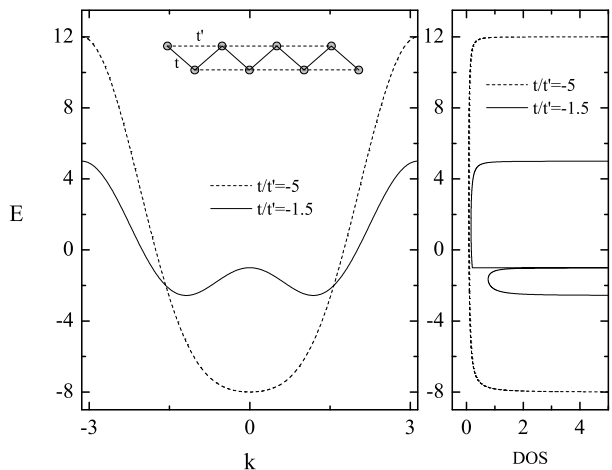} \caption{Dispersion carves
(left) and density of states (right) of pure chain for $t=-5, t'=1$
(dashed) and $t=-1.5, t'=1$ (solid).}\label{Fig1}
\end{figure}

In more realistic representation of the problem, hopping to the
next-nearest neighbors should be taken into account. In this paper
we want to address the stability of the anomalies, against the
introduction of hopping to next neighbors. We restrict ourselves to
hopping to the second neighbor only. Generalization for other
neighbors is straightforward. By doing so, the results will be
applicable to probing, recently proposed \cite{Rodriguez}
delocalization transition in low dimensional systems with long-range
hopping. We consider the one-dimensional Anderson model
\begin{equation}\label{model}
    t'(\Psi_{n+2}+\Psi_{n-2})+t(\Psi_{n+1}+\Psi_{n-1})+\epsilon
    U_n\Psi_n=E\Psi_n
\end{equation}
with nnn hopping and weak disordered potential $\epsilon U_n$, where
$\langle U_n\rangle=0$ and $\langle
U_nU_m\rangle=\sigma^2\delta_{nm}$. This model also can be viewed as
a system of two coupled chains in the way that is illustrated in
Fig. \ref{Fig1}. It is studied extensively \cite{Japaridze07} as an
extension of one-dimensional Hubbard model, called $t-t'$ Hubbard
chain, and exhibits a rich phase diagram. Pure chain has a
dispersion relation which is quadratic in $\cos k$
\begin{equation}\label{dispersion}
    E(k)=2t\cos k+2t'\cos 2k, \ \  -\pi<k<\pi,
\end{equation}
where unit lattice spacing is assumed. We will only consider
positive $t'$ on account of symmetry. Dispersion curves for two
different ratios of $|\frac{t}{t'}|>4$ and $|\frac{t}{t'}|<4$ are
plotted in Fig. \ref{Fig1}. In the latter case, there are two pairs
of wave vectors, carrying the same energy at the bottom of the band.
In other word, there are two propagating channels for that part of
spectrum. For each case the density of states (DOS) is shown in the
right panel. Additional singularity of DOS inside the band for
$|\frac{t}{t'}|<4$ is an internal band edge corresponding to new
channel. As $|\frac{t}{t'}|$ increases, this singularity moves
toward the bottom of the band and disappears at $|\frac{t}{t'}|=4$,
after which there will be single channel at entire band.

\section{Transfer Matrix}

Propagation along the chain according to Eq. (\ref{model}) can be
described by using $4\times 4$ transfer matrices,
\begin{equation}\label{TM}
    \left(\begin{array}{l}
       \Psi_{n+2} \\
       \Psi_{n+1} \\
       \Psi_{n} \\
       \Psi_{n-1}
     \end{array}\right)=
     \left(\begin{array}{cccc}
             -\frac{t}{t'} & \frac{E-\epsilon U_n}{t'} & -\frac{t}{t'} & -1 \\
             1 & 0 & 0 & 0 \\
             0 & 1 & 0 & 0 \\
             0 & 0 & 1 & 0
           \end{array}
     \right)
     \left(\begin{array}{l}
       \Psi_{n+1} \\
       \Psi_{n} \\
       \Psi_{n-1} \\
       \Psi_{n-2}
     \end{array}\right).
\end{equation}
The transmission channels of pure system can be distinguished in
terms of eigenvalues and corresponding eigenvectors of transfer
matrix for $\epsilon=0$. We have four eigenvalues which appear in
pairs $\lambda_i,\lambda_i^{-1}$. Each pair corresponds to a
right-going and a left-going plane waves. It should be noted that
each mode is propagating as long as corresponding eigenvalue has
unit norm and is evanescent otherwise. We will return to this point
and will give the range of energy for each channel.

In a similar manner, by addition of random potential one obtains
pairs of LEs $\pm\gamma_i,i=1,2$ for product of transfer matrices.
Each LE has a contribution to conductance but regarding the
eigenstates, for which $k$ is not a good quantum number in presence
of disorder, the smaller LE gives the localization length. The
transfer matrix in Eq. (\ref{TM}) will be used for numerical
calculation of LEs.

\section{Perturbation theory}

As usual we define the variables $R_n=\frac{\Psi_{n+1}}{\Psi_n}$ and
rewrite Eq. (\ref{model}) in terms of them. Also it is appropriate
to scale energies with $t'$ such that $\frac{t}{t'}=h$,
$\frac{U_n}{t'}\rightarrow U_n$ and $\frac{E}{t'}\rightarrow E$. We
have
\begin{equation}\label{model-R}
    R_{n+1}R_n+\frac{1}{R_{n-1}
    R_{n-2}}+h(R_{n}+\frac{1}{R_{n-1}})=E-\epsilon U_n.
\end{equation}
Following the reference \onlinecite{Itzykson} we use the ansatz
\begin{equation}\label{ansatz}
    R_n=a e^{B_n \epsilon + C_n \epsilon^2 + \cdots},
\end{equation}
and by inserting in Eq. (\ref{model-R}) and collecting terms of same
order in $\epsilon$ we obtain recursive equations for
$a,B_n,C_n,\ldots$. Up to second order in $\epsilon$ we have
%
%\begin{widetext}
\begin{subequations}\label{aBC}
\begin{eqnarray}
  && a^2+\frac{1}{a^2}+h(a+\frac{1}{a}) = E, \label{eq-a}\\
  && a^2(B_{n+1}+B_n)-\frac{1}{a^2}(B_{n-1}+B_{n-2}) \nonumber \\
  && \hspace{2.5cm} +h(aB_n-\frac{1}{a}B_{n-1})=-U_n, \label{eq-B}\\
  && a^2\left[C_{n+1}+C_n+\frac{1}{2}(B_{n+1}+B_n)^2\right]\nonumber \\
  && \hspace{1cm} +\frac{1}{a^2}\left[-C_{n-1}-C_{n-2}+\frac{1}{2}(B_{n-1}+B_{n-2})^2\right]\nonumber\\
  && \hspace{1cm} +h\left[a(C_n+\frac{1}{2}B_n^2)+\frac{1}{a}(-C_{n-1}+\frac{1}{2}B_{n-1}^2)\right]=0.\nonumber \\ \label{eq-C}
\end{eqnarray}
\end{subequations}
%\end{widetext}
%
The LE is given by
\begin{equation}\label{LE-def}
    \gamma(E)=\lim_{N\rightarrow \infty}\frac{1}{N}\sum_{n=1}^N \log R_n.
\end{equation}
Using Eqs. (\ref{ansatz}) and (\ref{LE-def}) an expansion for LE can
be obtained as follows
\begin{equation}\label{LE-exp}
    \gamma(E)=\log a +\epsilon \langle B\rangle+\epsilon^2\langle C\rangle+\cdots.
\end{equation}
Angular brackets denote the ensemble average. In order to calculate
this averages we take average of both sides of Eqs. (\ref{eq-B}) and
(\ref{eq-C}). Then we get
\begin{eqnarray}
  \langle B\rangle &=& 0, \\ \label{mean}
  \langle C\rangle &=& -\frac{(a^2+\frac{1}{a^2})[\rho(1)+\rho(0)]+\frac{h}{2}(a+\frac{1}{a})\rho(0)}
  {2(a^2-\frac{1}{a^2})+h(a-\frac{1}{a})},\ \ \ \
\end{eqnarray}
where $\rho(\tau)$ is the autocovariance function $\langle
B_{n+\tau}B_n\rangle$. Recursive equation (\ref{eq-B}) is an
autoregressive process of third order. Covariances $\rho(\tau)$ can
be determined by the following set of Yule-Walker equations for
$B$-process
\begin{subequations}
\begin{eqnarray}
  \rho(0) &=& \phi_1 \rho(1) + \phi_2 \rho(2) + \phi_3 \rho(3) + \frac{\sigma^2}{a^4},\\
  \rho(1) &=& \phi_1 \rho(0) + \phi_2 \rho(1) + \phi_3 \rho(2), \\
  \rho(2) &=& \phi_1 \rho(1) + \phi_2 \rho(0) + \phi_3 \rho(1), \\
  \rho(3) &=& \phi_1 \rho(2) + \phi_2 \rho(1) + \phi_3 \rho(0),
\end{eqnarray}
\end{subequations}
where $\phi_1=-(1+\frac{h}{a})$,
$\phi_2=\frac{1}{a^4}+\frac{h}{a^3}$ and $\phi_3=\frac{1}{a^4}$.

By solving this set of equations we find
\begin{subequations}\label{covariance}
\begin{eqnarray}
  \rho(0) &=& \frac{\sigma^2}{a^4M} (-1+\phi_2+\phi_1\phi_3+\phi_3^2), \label{covariance-a}\\
  \rho(1) &=& -\frac{\sigma^2}{a^4M} (\phi_1+\phi_2\phi_3), \label{covariance-b}\\
  \rho(2) &=& -\frac{\sigma^2}{a^4M} (-\phi_2+\phi_1^2-\phi_2^2+\phi_1\phi_3), \\
  \rho(3) &=& -\frac{\sigma^2}{a^4M}
  (\phi_3+2\phi_1\phi_2-\phi_2\phi_3-\phi_1\phi_2^2-\phi_1\phi_3^2 \nonumber\\
  && \ \ \ \ \ \ \ \ \ \ \ \ \ \ \ \ +\phi_1^2\phi_3+\phi_2^2\phi_3+\phi_1^3-\phi_3^3),
\end{eqnarray}
\end{subequations}
with $M=(1+\phi_1-\phi_2+\phi_3) (-1+\phi_1+\phi_2+\phi_3)
(1+\phi_2+ \phi_3(\phi_1-\phi_3))$. By inserting Eqs.
(\ref{covariance-a}) and (\ref{covariance-b}) in Eq. (\ref{mean})
and using Eq. (\ref{LE-exp}) we obtain the LE up to second order (in
the original energy scale)
\begin{equation}\label{LE}
   \frac{1}{\xi}=\mathfrak{Re}(\gamma)=\frac{-\sigma^2\epsilon^2}{2\left[
   2t'(a^2-\frac{1}{a^2})+t(a-\frac{1}{a})\right]^2}.
\end{equation}
Note that inside the band, $a$ is pure phase. This expression
reduces to the well-known result for Anderson model at the limit
$t'\rightarrow 0$. All that remains is to find roots of Eq.
(\ref{eq-a}). Equation (\ref{eq-a}) is in fact the characteristic
equation for eigenvalues of transfer matrix of pure system. Inside
the energy band where we have $a=e^{i k}$, it is nothing but the
dispersion relation in Eq. (\ref{dispersion}), from which we have
\begin{eqnarray}
\cos k_{\pm}=\frac{1}{4}(-h\pm\sqrt{h^2+4E+8}).
\end{eqnarray}
Depending on the sign of $h$, one of branches produces the energy
band $-2-\frac{h^2}{4}<E_1<2+2|h|$ and the other
$-2-\frac{h^2}{4}<E_2<2-2|h|$, for $|h|<4$. We will call them first
and second channel respectively. For $|h|>4$, second channel gets
closed and the first one spans the interval $2-2|h|<E_1<2+2|h|$.

\section{Poles of $\rho(\tau)$ and anomalies}

Perturbation expansion in Eq. (\ref{LE-exp}), diverges in different
orders for special energies. As we mentioned, this signals an
anomaly and the order of divergent term is related to the order of
anomaly. As it is shown for Anderson model, first divergence is
showed up in the fourth order at the middle of the band which
corresponds to the principal anomaly (KW anomaly). The expansion is
finite up to the second order for the model studied here as well.

\begin{figure}[t]
\subfigure{\epsfxsize8truecm \epsffile{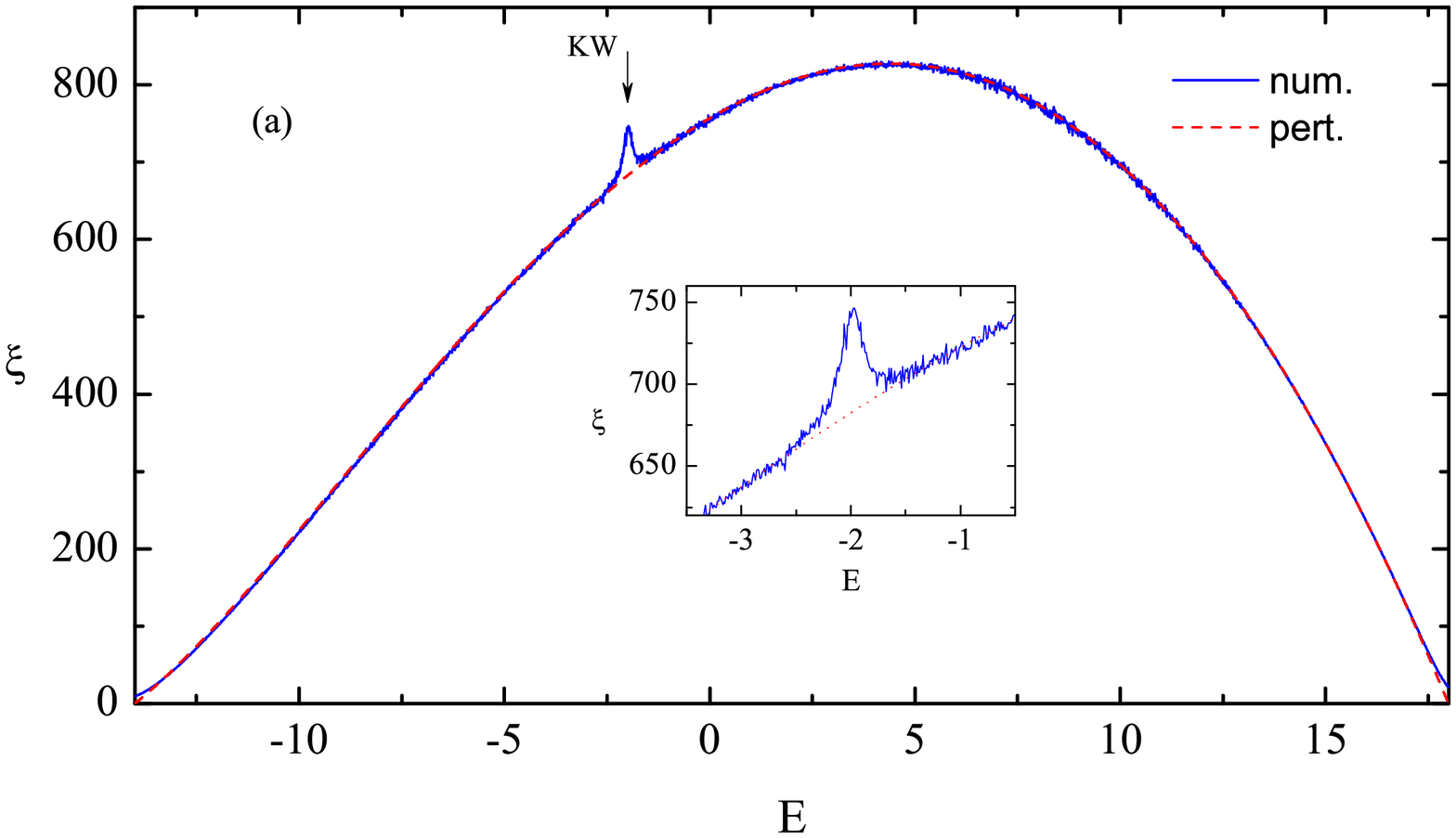}\label{Fig2a}}
\subfigure{\epsfxsize8truecm \epsffile{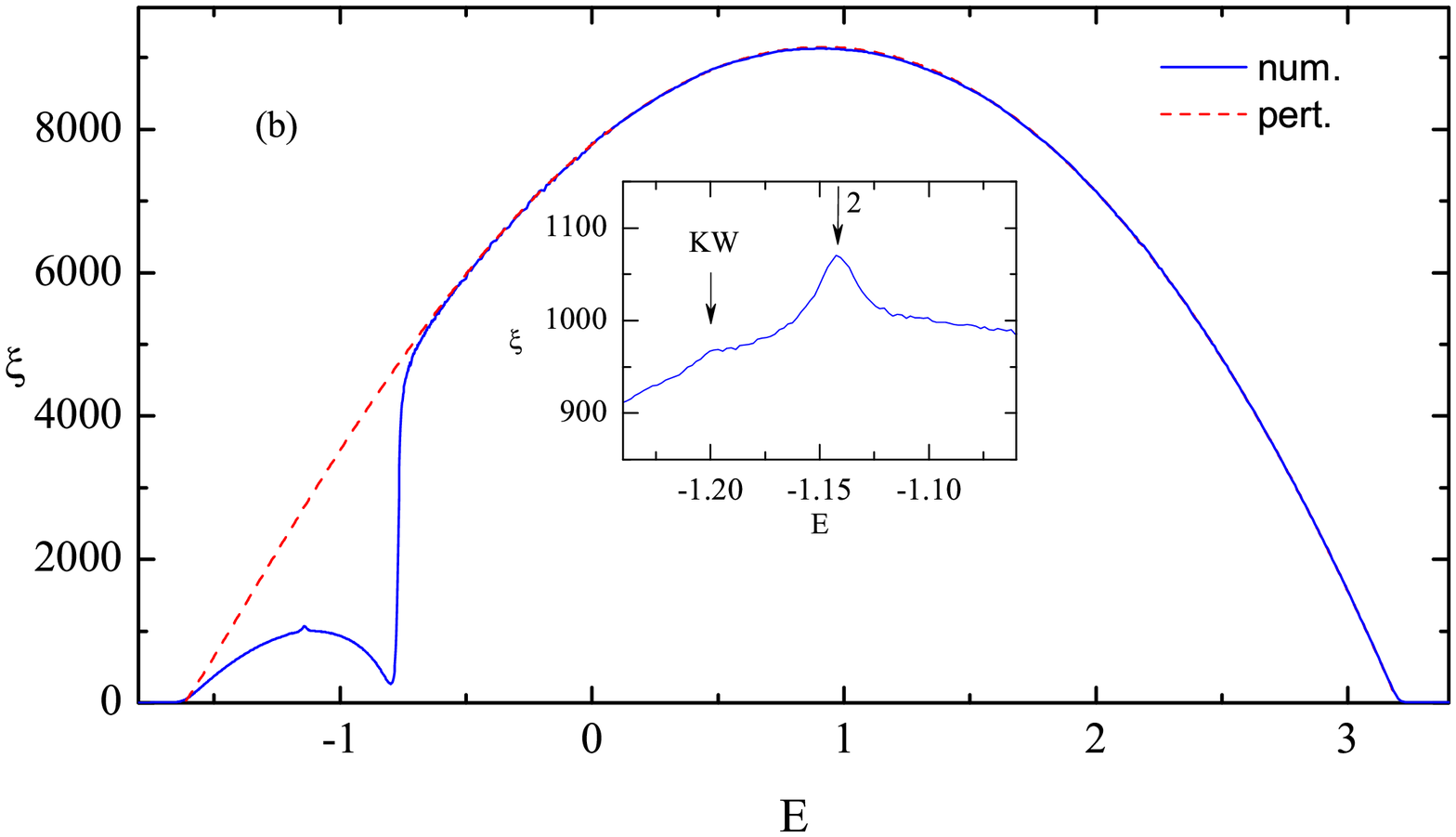} \label{Fig2b}}
\subfigure{\epsfxsize8truecm \epsffile{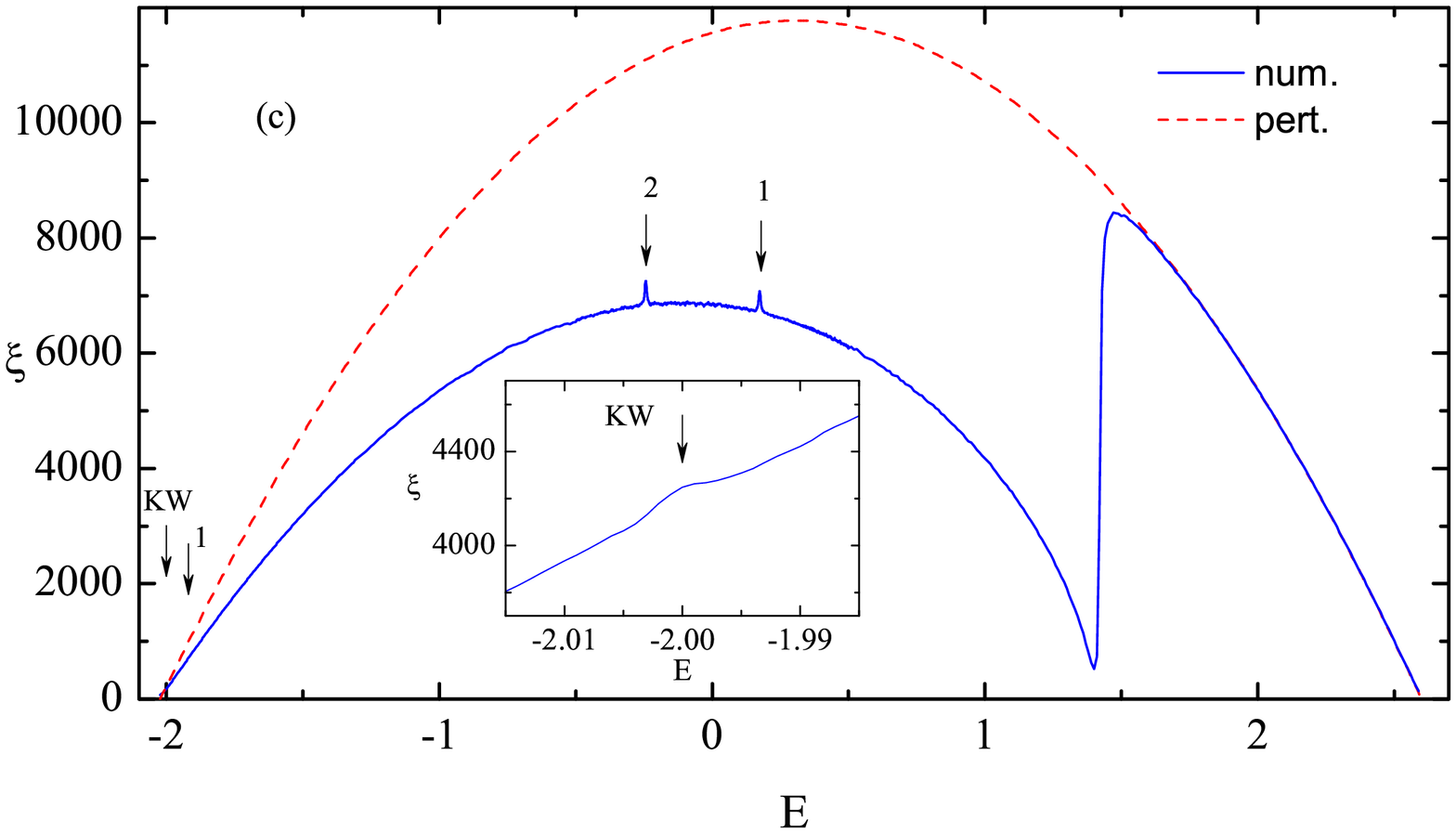}\label{Fig2c}}
\caption{(Color online) Localization length obtained from numerical
calculation via transfer matrix method (solid) and analytical
perturbation theory (dashed). White noise disorder with uniform
distribution and zero mean is used. Arrows indicate the anomalies
and their positions are determined from poles of covariance function
$\rho(\tau)$. Numbers denote the channel in which the poles show up.
(a) $t=8,t'=1,\sigma^2=\frac{3}{4}$, inset is a closer view of
anomaly, (b) $t=1,t'=0.6,\sigma^2=\frac{1}{300}$, inset is a closer
view of anomalies, and (c) $t=0.3,t'=1,\sigma^2=\frac{1}{300}$,
inset: $t=0.8,t'=1,\sigma^2=\frac{1}{1200}$. }\label{Fig2}
\end{figure}

The point that has not been noticed is the appearance of divergences
at the level of covariance functions $\rho(\tau)$. Here we show that
covariance functions $\rho(\tau)$ possess some poles on real energy
axis, yet the final result in Eq. (\ref{LE}) is finite. By exact
numerical calculation we show that the localization length enhances
at these poles. Depending on $h$, there are four situations as
follows. Without loss of generality let us consider positive $h$
hereafter.

$\mathbf{h>4}$. By inserting a small hopping term, $t'$, in the
hamiltonian, results will smoothly deviate from that of ordinary
Anderson model. It will cause an asymmetry in the localization
length vs. energy about the zero energy. Other surprising fact is
that the anomaly at the band center survives and shifts from the
center [Fig. \ref{Fig2a}]. If we suppose that it would occur at the
same fraction of wavelength to the lattice spacing, which was seen
for Anderson model, we can estimate the energy by using dispersion
relation in Eq. (\ref{dispersion}) for $k=\frac{\pi}{2}$, which
gives $E=-2t'$. Now by looking at $\rho(\tau)$ we can see that there
is one pole at $E=-2$ (root of the term $1+\phi_1-\phi_2+\phi_3$ in
$M$ for first channel) which is the same energy, $E=-2t'$, in the
original energy scale. This pole indeed corresponds to the KW
anomaly which now appears away from the center. The anomalous
behavior of localization length and deviation from parturbative
result at this point can be seen clearly in Fig. \ref{Fig2a}. It is
also present in other cases but appears weaker. The second channel
is evanescent in this case.

$\mathbf{\frac{4\sqrt{6}}{9}<h<4}$. Third factor in denominator,
$1+\phi_2+\phi_3(\phi_1-\phi_3)$, has other roots which satisfy the
cubic equation $x^3=3px-2q$, where $p=\frac{2}{3}$, $q=\frac{h}{2}$
and $x=(a+\frac{1}{a})=2\cos k_{\pm}$. Positive and negative signs
correspond to first and second channels, respectively. In this case
which we have $q^2-p^3>0$, the equation has one real root
$x=-(q+\sqrt{q^2-p^3})^{1/3}-(q-\sqrt{q^2-p^3})^{1/3}$ which gives a
valid energy in second channel only. In the first channel we have
one pole so far [Fig. \ref{Fig2b}].

$\mathbf{h=\frac{4\sqrt{6}}{9}}$. In this case ($q^2-p^3=0$), the
cubic equation has three real roots, one of which has multiplicity
2, $x_1=-2q^{1/3}, x_2=x_3=q^{1/3}$. $x_1$ is actually the root in
previous case and corresponds to the second channel. The other root
gives rise a pole of second order in the first channel.

\begin{figure}[t]
\subfigure{\epsfxsize8truecm \epsffile{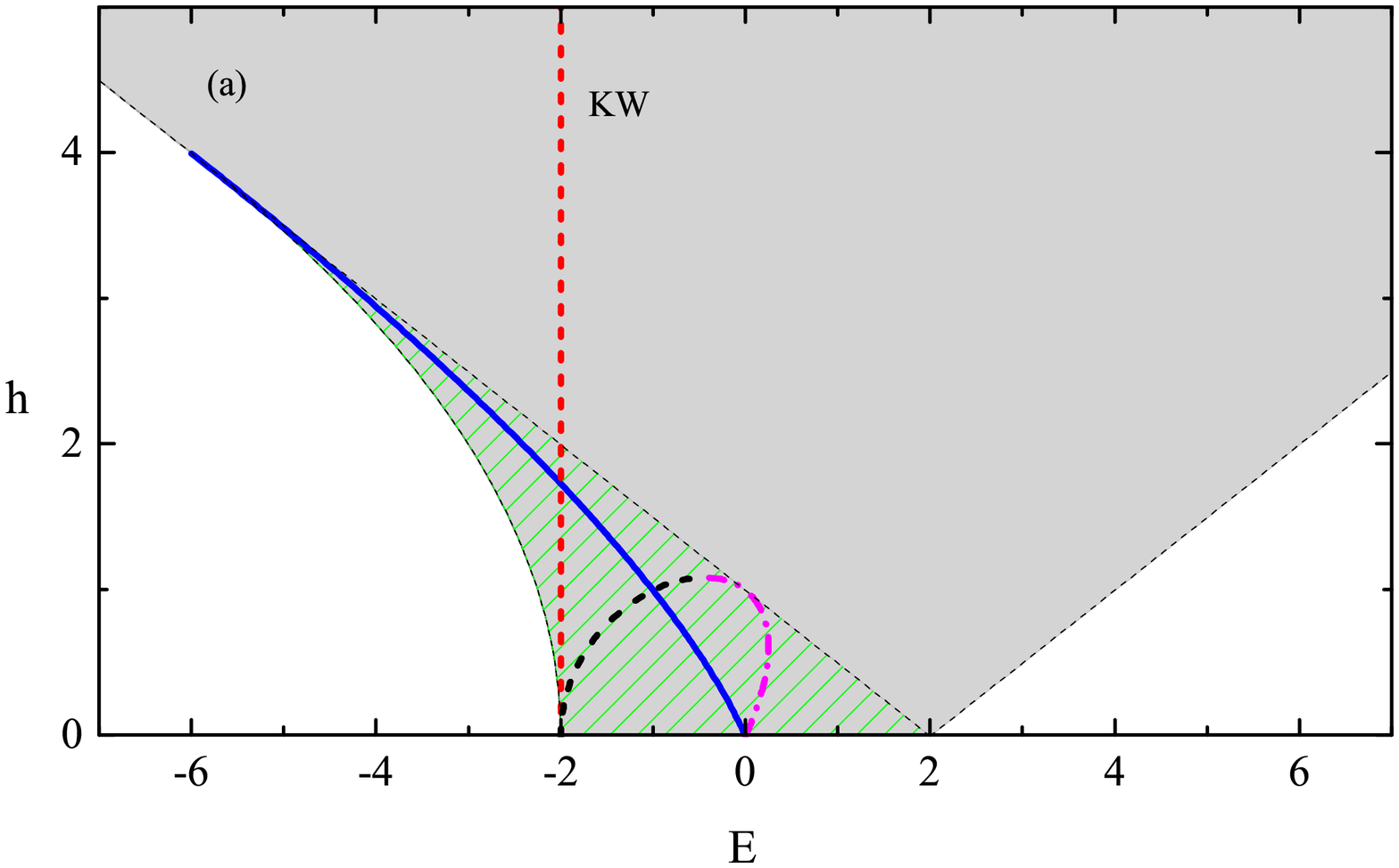}\label{Fig3a}}
\subfigure{\epsfxsize8truecm \epsffile{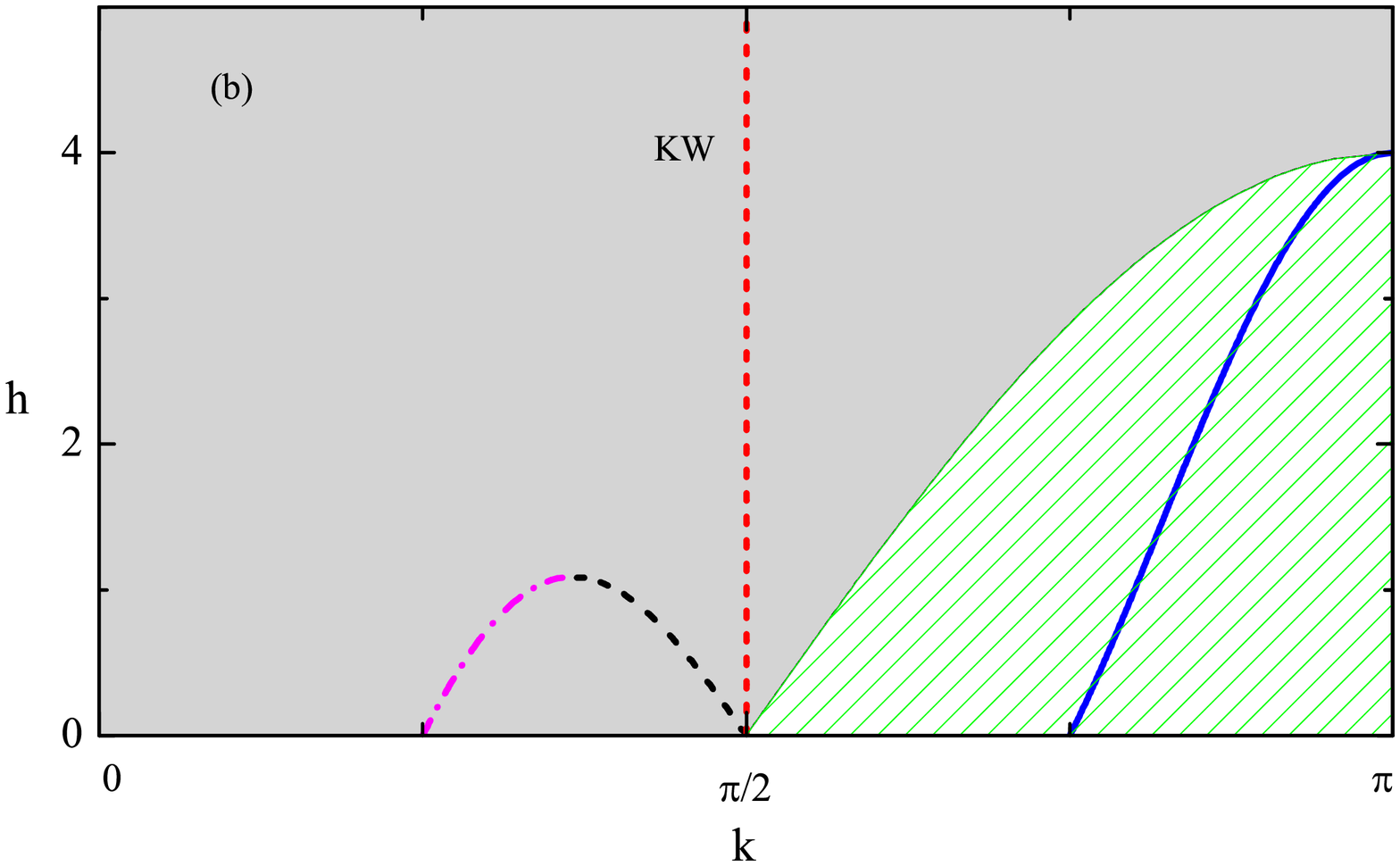} \label{Fig3b}}
\caption{(Color online) Transmission channels of pure system and
trajectory of poles of $\rho(\tau)$. Two channels are specified with
gray color (first channel) and green shade (second channel). As it
is clear from (b), red dashed (KW), black dashed and dashed-dotted
lines correspond to first channel and solid blue line corresponds to
the second channel. Thin dashed lines at the band edges in (a) are
also nonanalytic poles of $\rho(\tau)$. }\label{Fig3}
\end{figure}

$\mathbf{h<\frac{4\sqrt{6}}{9}}$. We have three distinct real roots
in this case ($q^2-p^3<0$), which can be expressed in the standard
form,
$x_1=2\sqrt{p}\cos(u/3),x_2=2\sqrt{p}\cos(u/3+2\pi/3),x_3=2\sqrt{p}\cos(u/3+4\pi/3)$,
where $\cos u=-q/(p\sqrt{p}),0<u<\pi$. Again one of them is in
second channel and other two are in the first channel [Fig.
\ref{Fig2c}]. One of the later poles seems to be absent in numerical
results and needs to be discussed beyond the second-order
perturbation. These cases are summarized in the Fig. \ref{Fig3}.

The wave vector of last three poles is given by
\begin{equation}
    k=\arccos\frac{x}{2}.
\end{equation}
Due to the dependence of $x$ on $h$, wave vector changes
continuously by varying the ratio of hopping integrals [Fig.
\ref{Fig3b}]. Thus the value of wavelength at these anomalies will
not be necessarily rational.

We shall mention two other essential features in Fig. \ref{Fig2}.
(\textit{i}) Apart from in the anomalies, perturbative result
deviates significantly from numerical result for those energies at
which two channels are open [Fig. \ref{Fig2b} and \ref{Fig2c}].
Analytical result based on Eq. (\ref{ansatz}) presents the
perturbation around the solutions with single wave vector while the
numerical method produces a mixture of two solutions. The fact that
the anomalies are obtained correctly by perturbation theory,
supports the above statement.

It is worth mentioning the limit of zero $h$ where the whole
spectrum is degenerate. Covariances in Eq. (\ref{covariance}) are
divergent, however LE in Eq. (\ref{LE}) has a well-defined limit. We
have nnn hopping ($t'$) only at this limit and the system transforms
to two decoupled chains with nearest-neighbor hopping. So one
expects the result of ordinary Anderson model ($\xi_A$) with doubled
lattice constant, i.e.,  $\xi\rightarrow2\xi_A$. But the factor is
$4$ rather than $2$. This suggests to use decoupled chains as a
starting point to develop the perturbation theory.

(\textit{ii}) The sudden change in localization length happens at
the internal band edge and is in coincidence with van-Hove
singularity in DOS of pure system.

\section{conclusion}

In conclusion, we show that KW anomaly exists in presence of nnn
hopping, where we have lower symmetry of hamiltonian for vanishing
disorder (broken particle-hole symmetry), and occurs at the same
wavelength as in the Anderson model. We also demonstrate that the
anomaly could be identified by certain divergences at first order of
perturbation theory. Three other singularities turn out to exist
which are not attributed to single wavelength and may even
correspond to incommensurate ratio of wavelength to the lattice
spacing which is in striking contrast to the known anomalies in
Anderson model.

\begin{thebibliography}{99}
\bibitem{PRM} A. Crisanti, G. Paladin and A. Vulpiani \textit{Products of Random Matrices in Statistical
Physics} (Springer-Verlag, Berlin, Heidelberg, 1991).
%
\bibitem{Thouless} D.J. Thouless, in Ill-Condensed Matter, Les Houches Summer School,
1978, edited by R. Balian, R. Maynard, G. Toulouse (North-Holland,
New York, 1979).
%
\bibitem{MacK81} G. Czycholl, B. Kramer and A. MacKinnon, Z. Phys. B - Condensed
Matter \textbf{43}, 5 (1981).
%
\bibitem{Kappus} M. Kappus and F. Wegner, \textit{Z. Phys.} \textbf{B} \textbf{45}, 15–21 (1981).
%
\bibitem{Lambert84} C.J. Lambert, J. Phys. C: Solid State Phys. \textbf{17}, 2401-2414 (1984) .
%
\bibitem{Hayn88}R. Hayn and W. John,  \textit{Z. Phys.} \textbf{B} \textbf{70}, 331-340 (1988).
%
\bibitem{Schomerus03} H. Schomerus and M. Titov, Phys.
Rev. B. \textbf{67}, 100201(R) (2003).
%
\bibitem{Titov05} M. Titov and H. Schomerus, Phys. Rev. Lett, \textbf{95}, 126602 (2005).
%
\bibitem{Anderson}P.W. Anderson, D. J. Thouless, E. Abrahams, and D. S.
Fisher, Phys. Rev. B \textbf{22}, 3519 (1980).
%
\bibitem{DG} B. Derrida and E. Gardner, J. Phys. (Paris) \textbf{45}, 1283
(1984).
%
\bibitem{Schulz} H. Schulz-Baldes, \textit{Operator Theory:
Advances and Applications}, \textbf{174}, 159–172 (Birkhauser,
2007).
%
\bibitem{Yudson} V.I. Yudson and V.E. Kravtsov, AIP Conf. Proc., \textbf{1134}, 31-35 (2009).
%
\bibitem{Rodriguez}A. Rodriguez, V. A. Malyshev, G. Sierra, M. A. Martin-Delgado, J. Rodriguez-Laguna, and F.
Dominguez-Adame, Phys. Rev. Lett, \textbf{90}, 027404 (2003).
%
\bibitem{Japaridze07}See G. I. Japaridze, R. M. Noack, D. Baeriswy and L. Tincani, Phys. Rev. B \textbf{76}, 115118 (2007) and refrences therein.
%
\bibitem{Itzykson}E.J. Gardner, C. Itzykson and B. Derrida, J. Phys. A: Math. Gen. \textbf{17}, 1093 (1984).
%
\end{thebibliography}
\end{document}